\documentclass[]{article}
\usepackage{graphicx}

\title{Big Data Challenges in Genome Informatics}
\author{Ka-Chun Wong}

\begin{document}

\maketitle

\begin{abstract}
In recent years, we have witnessed a dramatic data explosion in genomics, thanks to the improvement in sequencing technologies and the drastically decreasing costs. We are entering the era of millions of available genomes. Notably, each genome can be composed of billions of nucleotides stored as plain text files in GigaBytes (GBs). It is undeniable that those genome data impose unprecedented data challenges for us. In this article, we briefly discuss the big data challenges associated with genomics in recent years. 
\end{abstract}

\section*{Introduction}

Since 1990s, the whole genomes of different species have been sequenced by their corresponding genome sequencing projects. In 1995, the first free-living organism Haemophilus influenzae was sequenced by the Institute for Genomic Research. In 1996, the first eukaryotic genome (Saccharomyces cerevisiase) was completely sequenced. In 2000, the first plant genome, Arabidopsis thaliana, was also sequenced by Arabidopsis Genome Initiative. In 2003, the Human Genome Project (HGP) announced its completion. Following the HGP, the Encyclopedia of DNA Elements (ENCODE) project was started, revealing massive functional elements on the human genome in 2011 \cite{encode2004encode}. The drastically decreasing cost of sequencing also enables the 1000 Genomes Project and Roadmap Epigenomics Project to be carried out. Their results have been published in 2012 and 2015 respectively \cite{10002010map,kundaje2015integrative}. Nonetheless, the massive genomic data generated by those projects impose an unforeseen challenge for big data analysis at the scale of GigaBytes (GBs) or even TeraBytes (TBs). 

In particular, Next-Generation Sequencing (NGS) technologies have enabled massive data generation for different genomes \cite{wong2014snpdryad,mardis2008impact}; for instance, DNA sequencing, protein-DNA binding occupancy \cite{wong2013dna} (e.g. ChIP-seq \cite{visel2009chip}, ChIP-exo \cite{rhee2011comprehensive}, and  ChIA-PET \cite{fullwood2009oestrogen}), bisulfite sequencing \cite{bock2005biq}, transcriptome sequencing (e.g. RNA-seq \cite{mortazavi2008mapping}), and chromatin interaction sequencing (e.g. Hi-C \cite{lieberman2009comprehensive}). Thanks to the relatively low costs, those NGS technologies have been readily applied to human genomes nowadays. The international projects aforementioned have been successfully launched, leading to massive NGS data accumulation at an unprecedentedly fast pace. Nonetheless, current integrative analyses are usually limited to traditional machine learning and data mining methods such as pair-wise correlation analysis, statistical tests, classification, and feature extraction \cite{wongprobabilistic}. Those methods are intentionally designed to generally fit into different types of data. However, the data from NGS is unique and different from the traditional data; for instance, the ChIP-seq data is sparse, noisy, and discontinuous. Special care has to be taken to alleviate and transform those challenges to be taken advantages of \cite{pmid25192742}. In addition, the NGS data is huge (in GigaBytes per each dataset) which imposes a difficulty in applying some of the existing statistical/computational methods. 

Therefore, different genome-scale problems been defined and being tackled to harness those genomic data. In this review, we aim at providing a concise literature review on those challenges.

\section*{Challenges}
\begin{figure}[h!]
	\includegraphics[width=\textwidth]{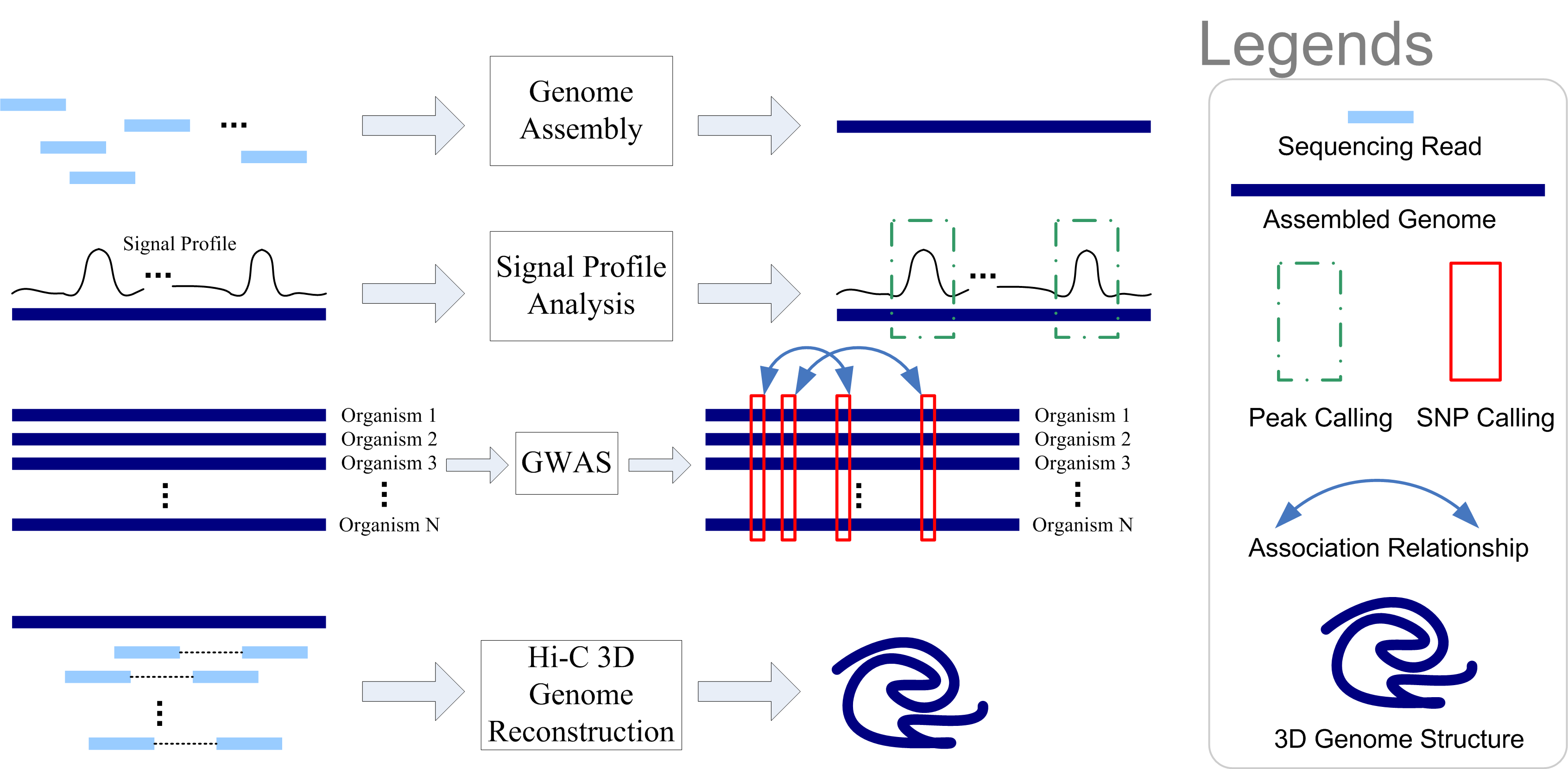}
	\caption{Big Data Challenges in Genome Informatics. The challenges are listed from top to bottom; namely, genome assembly, signal profile analysis, genome-wide association study (GWAS), and 3D genome structure reconstruction.}
	\label{fig:overview}
\end{figure}

In this section, we provide a concise review on the big data challenges which are individually described in the following subsections. Especially, the challenges have been visualized and depicted in Figure \ref{fig:overview} for reader-friendliness.

\subsection*{De Novo Genome Assembly}
The advancement in DNA sequencing technologies has enabled the assembly of whole genome in an economical and fairly accurate way \cite{pmid21307932}. Nonetheless, a genome cannot be easily identified in one piece from wet-labs. Limited by our current DNA sequencing technologies, each genome has to be shattered into different small pieces (short DNA sequences or reads) before their DNA nucleotides can get sequenced and identified as shown in Figure \ref{fig:overview}. Therefore, we come to the \textit{de novo} genome assembly problem: to sequence and identify a genome, we have to 'stitch' those short DNA sequences into a single and consistent DNA genome. There are different benchmark measurements such as N50, total length, and number of missing nucleotides. If we already have a reference genome, then the measurements can be more solid than the previous measures such as NG50 and genome fraction. If reference genome annotation is available, the number of genes covered can be a good measurement. More details can be found in \cite{pmid23422339}. To solve this kind of genome assembly problems (in GBs or TBs), there are many computational methods proposed in the past. Nonetheless, most of them depend on the construction of \textit{de brujin} graph which is memory-consuming and computationally intensive. According to the recent benchmark study, different genome assembly methods show result disagreement with each other \cite{pmid23870653}. In addition, the sequencing errors incurred by wet-lab experimental conditions are unavoidable, making the genome assembly problem even more complicated than we have imagined \cite{pmid21307932}. Therefore, the genome assembly problem remains as a big data challenge to be solved.

\subsection*{Genome Signal Profile Analysis}
In addition to genome assembly, there are different genome-wide signals such as gene regulation (e.g. protein-DNA binding interactions) and epigenetic interactions (e.g. DNA methylation) as shown in Figure \ref{fig:overview}. Therefore, it is essential for us to look into those information. To this end, different genome-wide biotechnologies have been developed such as ChIP-seq, DNase-seq, RNA-seq, CLIP-seq, DNA methylation array assay, Bisulphite sequencing, Repli-seq, and CAGE. To gains insights into those data, tremendous efforts have been made to pre-process the data such as read trimming \cite{pmid24695404}, sequencing error correction \cite{pmid22492192}, sequencing replicates \cite{pmid24322726}, and read mapping \cite{pmid21278192}. After the data has been processed, downstream analysis methods can be applied to reveal genome-wide signals from it; for instance, multiple signal profile integrative analysis \cite{pmid25192742,wongprobabilistic} and signal profile peak calling \cite{zhang2008model}. In particular, the multiple signal profile analysis is very important for us to understand the complex behavior of the genome-wide signals \cite{pmid25192742}. Unfortunately, each signal profile is proportional to genome size since it has a genome-wide coverage (usually in GBs). Therefore, if we have multiple signal profiles (e.g. hundreds from the ENCODE consortium), the computational scalability issue has to be taken into serious account. Another issue is that the past wet-lab studies are very limited to fine-scale knowledge (e.g. single gene study). Therefore, the genome-wide result verification is very difficult to be carried out. At the current stage, we heavily rely on null hypothesis testing to ascertain the results' statistical significances. Therefore, we can foresee that the genome signal profile analysis will still be a big data challenge in genome informatics.

\subsection*{Multiple Genomes: GWAS}
Single Nucleotide Polymorphisms (SNPs) are believed to be play major roles in different diseases \cite{10002010map}. To call for a SNP, since each SNP stems from natural selection which is acting and 'fixing' the SNP to favor genetic adaption \cite{barreiro2008natural}, we need multiple genomes for comparison (as shown in Figure \ref{fig:overview}) and satisfy certain statistical significance criteria \cite{altmann2012beginners}; for instance, the simplest criteria is how often an allele occurs across a given genome population. However, there is not any single golden measurement for such a task. Specifically, one of the major challenges in SNP calling is to handle the massive genome-wide data incurred from high-throughput sequencing. To compare multiple genomes for SNP calling, we have to align the genomes first. Although there exists an algorithm for mathematically optimal genome alignment (i.e. Smith-Waterman Alignment), its computational complexity is exponential to the genome size, and thus practically infeasible. Therefore, different heuristic genome alignment methods have been proposed. The most successful example is MUMmer which proposes an anchor-based alignment methodology \cite{kurtz2004versatile}. VISTA has also been proposed for multiple genome alignment visualization \cite{frazer2004vista}. However, most of the previous methods were initially designed and targeted for short genomes such as prokaryotic genomes;  new computational methods have to be developed, given that we are going to have thousands of eukaryotic genomes available; for instance, the 1000 human genomes from the 1000 Genomes Project \cite{10002010map}.

Another challenge is that, given a multiple genome alignment, we have to scan each position for SNP calling as shown in Figure \ref{fig:overview}. Especially, it is usually the case that we wish to find the SNPs which are associated with certain phenotypes such as human diseases. Therefore, it is named genome-wide association study (GWAS). Given the existing genome sizes measured in GBs, GWAS studies are famous for its computational intensive nature. If we take into account the combinatorial nature of the association, the possible search space could be exponential to genome sizes. Therefore, GWAS study is also considered as a big data problem in genomics.

\subsection*{3D Genome Structure Reconstruction}
In recent years, Hi-C technology has been developed and applied to reveal the three dimensional organizations of different cell lines by the chromosome conformation capture method \cite{pmid22652625}. In particular, there is increasing evidence that long-range chromatin interactions are related to gene co-expression \cite{pmid25965262,pmid24141950} as well as protein-DNA interactions \cite{pmid22675074,pmid25938943}. Therefore, it is essential to comprehensively identify and reconstruct the three-dimensional (3D) genome shape from those long-range chromatin interactions for understanding genomes in the three-dimensional space. Given the GB data size of genome as well as its three-dimensional nature, such a 3D genome reconstruction is doomed to be another big data challenge.

%
%
%
%
%
%
%
%
%
%
%
%


\section*{Discussion}

In this study, we have discussed several big data challenges in genome informatics. Especially, we envision that those challenges will be even intense in the near future, given the maturing and cost-effective sequencing technologies. Several future directions are deemed promising: (1) Third-generation sequencing technologies \cite{schadt2010window} have been developed and being refined to be of practical uses. Although its sequencing error rate is still high, we believe that those third-generation sequencing technologies would enable another wave of big data challenges in genome informatics. (2) Single cell sequencing is another promising direction. For now, we usually just focus on specific cell types or tissue types. In the future, our sequencing technologies would enable us to look at each of the individual cells which holds great potential to trigger the next levels of big data challenges. (3) Given the genome data in GBs or even TBs, high-performance computing frameworks such as MapReduce are definitely needed to handle the exponentially growing genome data in a scalable but still accurate manner.

\bibliographystyle{plain}
\bibliography{bmc_article} 

\end{document}